\documentclass[showpacs,preprintnumbers,amsmath,reprint,aps,pra,twocolumn]{revtex4-1}

\usepackage{bm}
\usepackage{graphicx} 
\usepackage{color}
\usepackage{amsmath}
\usepackage{feynmf}
\usepackage{slashed}

\begin{document}
 
\title{Two-photon total annihilation of molecular positronium}

\author{Jes\'{u}s P\'{e}rez-R\'{i}os}

\affiliation{Department of Physics and Astronomy, 
Purdue University,  47907 West Lafayette, IN, USA}

\author{Sherwin T. Love }

\affiliation{Department of Physics and Astronomy, 
Purdue University,  47907 West Lafayette, IN, USA}

\author{Chris H. Greene }

\affiliation{Department of Physics and Astronomy, 
Purdue University,  47907 West Lafayette, IN, USA}

\date{\today}

\begin{abstract}
The rate for complete two-photon annihilation of molecular positronium 
Ps$_{2}$ is reported. This decay channel involves a four-body 
collision among the fermions forming Ps$_{2}$, and two photons
 of 1.022 MeV, each, as the final state. The quantum 
 electrodynamics  result 
 for the rate of this process is found to be
  $\Gamma_{Ps_{2} \rightarrow \gamma\gamma}$ = 9.0 $\times 10^{-12}$ s$^{-1}$.
 This decay channel completes the most comprehensive decay chart for 
 Ps$_{2}$ up to date.
     
\end{abstract}

\maketitle

\section{Introduction}

Positronium or Ps, is the bound state of an electron 
and its antiparticle, the positron, forming a metastable 
hydrogen-like atom \cite{Positron-Physics}. In the 1940's, 
Wheeler speculated that two Ps atoms may form 
molecular positronium Ps$_{2}$, in analogy with two 
hydrogen atoms that can combine to form molecular hydrogen
\cite{Wheeler-1946}. In the same decade, calculations of 
the binding energy of Ps$_{2}$  were carried out, and it 
turned out to be 0.4 eV \cite{Hylleraas-1947}, supporting 
Wheeler's prediction. More recently, in 2007, 
Cassidy and Mills reported the first observation of
molecular positronium \cite{Cassidy-2007}. 

Molecular positronium can decay to different final states 
or channels. The characterization of the decay channels
 is essential in order to estimate its lifetime. Moreover, 
 the complete characterization of the Ps$_{2}$ decay channels and their
 partial widths could lead to the design of efficient detection 
 schemes for this molecule. For a bound state, the total 
 annihilation rate $\Gamma$ is determined as the sum of
 partial annihilation rates associated with each allowed 
 decay channel $\Gamma_{i}$, i.e., $\Gamma=\sum_{i=1,N}\Gamma_{i}$,
where $N$ represents the total number of decay channels. 
Each of the $\Gamma_{i}$ has to be computed by including
all the topologically distinct Feynman diagrams associated with such
channels, and in some cases, it can be important to include radiative
 corrections. Frolov has reported the most complete chart 
 of decay channels as well as partial annihilation rates 
 up to date \cite{Frolov-2009}, including all the main 
 decay channels, going from zero photon decay up to the 5-photon
 decay channel. However, a higher order decay channel of Ps$_{2}$ 
 involving two-photons as the final state 
  has not been considered in any estimation 
 of the Ps$_{2}$ lifetime, and apparently never previously contemplated as a 
 possible decay channel.

The present study reports the calculation of the  two-photon complete
annihilation rate of Ps$_{2}$, in which two electrons and two
 positrons annihilate simultaneously, producing two photons of 1.022 MeV energy
 each. The calculations have been carried out by using standard 
 techniques of quantum field theory, such as the Feynman rules 
 and trace technology \cite{Peskin}. This decay channel completes 
 the decay chart of Ps$_{2}$, previously reported in part by 
 Frolov \cite{Frolov-2009}, besides the six-photon and seven-photon decay 
 channels. While this decay is rare, it is worth mentioning that it provides a unique experimental signature of the presence of molecular positronium.
 
\section{Two-photon annihilation of Ps$_{2}$}

The annihilation of Ps$_{2}$ into two photons (denoted 
here as Ps$_{2} \rightarrow \gamma\gamma$) is governed by 
eight topological distinct Feynman diagrams. Four of them 
are shown in Fig.1. The rest of the diagrams emerge as cross 
terms of the ones shown in Fig. 1, {\it i.e.}, in which the momenta 
of the outgoing photons are interchanged. Fig. 1 shows that the decay 
channel at hand is a four-body event, where the energy-momentum vectors of 
the incoming fermions are labelled as $p_{1}$, $p_{2}$, $p_{3}$, 
and $p_{4}$, whereas the energy-momenta of the outgoing photons 
are labelled as $k_{1}$ and $k_{2}$. Here, the energy-momentum 
vectors are represented as $(E,\vec{p})$, and natural units 
($\hbar=c$ = 1, and $\alpha$ = 1/137, being fine structure constant)
 are assumed. 

The momenta of the electrons and positrons in Ps$_{2}$ are 
very low in comparison with their rest mass energy. Hence the binding energy of the Ps$_{2}$ molecule is negligible
in comparison with the rest mass energy of whatever of their
constituents. Therefore, the first non-vanishing term in the 
amplitude expansion can be obtained by substitution of 
the initial energy-momentum vectors by $(m_{e},0,0,0)$, 
instead of the initial energy-momentum vectors.  Here $m_{e}$ 
is the electron mass. Thus, in this approximation, the transition 
probability does not depend on the initial momenta $\vec{p}_{i}$. Within 
this approximation it is possible to establish a relationship 
between the annihilation rate and the probability to find 
the four fermions in the Ps$_{2}$ molecule to all be located at the
 same point in space. This information can be determined by generalizing the method
 employed for the calculation of the electron-positron 
 annihilation rate of Ps \cite{Peskin}, but 
going beyond the two-body perspective of that reference. This generalization 
leads to: 

\begin{eqnarray}
\label{eq-1}
\Gamma_{Ps_{2}\rightarrow \gamma \gamma}=\frac{|\Psi_{Ps_{2}}\left( 0,0,0,0 \right)|^{2}}{4}
\int \frac{d^{3}\vec{k}_{1}}{\left( 2\pi \right)^{3}2|\vec{k}_{1}|}
\frac{d^{3}\vec{k}_{2}}{\left( 2\pi \right)^{3}2|\vec{k}_{2}|}\nonumber \\
\left( 2\pi \right)^{4}\frac{\delta^{(4)}\left(p_{1}+p_{2}+p_{3}+p_{4}-k_{1}-k_{2}\right)}{\prod_{i=1,4}(2E_{i})}
|\mathcal{M}|^{2}. \nonumber \\ 
\end{eqnarray}

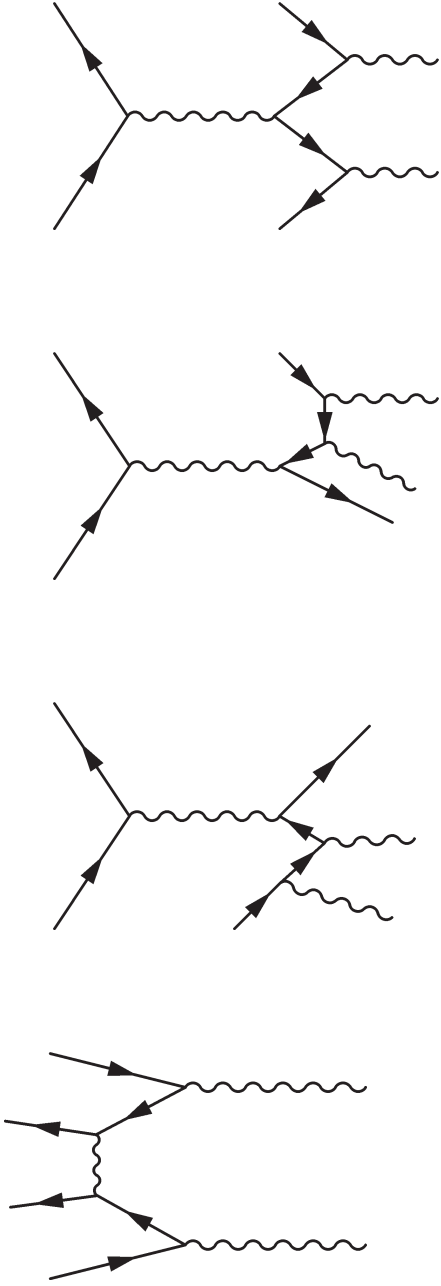
\begin{figure}[!ht]
\centering
\begin{minipage}{0.5\textwidth}
\begin{fmffile}{diagram1}
\begin{fmfgraph*}(60,30)
\fmfleftn{i}{2}\fmfrightn{o}{4}
\fmflabel{$e^-,p_1$}{i1}\fmflabel{$e^+,p_2$}{i2}
\fmflabel{$e^-,p_3$}{o1}\fmflabel{$e^+,p_4$}{o2}
\fmf{photon,label=$\gamma$}{v5,v2}
\fmf{photon,label=$k_{1}$}{o3,v3}
\fmf{photon,label=      $\  \ \ \ k_{2}$}{o4,v4}
\fmf{fermion}{i1,v5,i2}
\fmf{fermion}{v2,v3,o2}
\fmf{fermion}{o1,v4,v2}
\fmfforce{(0.75w,0.75h)}{v4}
\fmfforce{(0.6w,h)}{o1}
\fmfforce{(0.75w,0.25h)}{v3}
\fmfforce{(0.6w,0.0h)}{o2}
\fmfforce{(0.95w,0.25h)}{o3}
\fmfforce{(0.95w,0.75h)}{o4}
 \end{fmfgraph*}
 \end{fmffile}\quad\\
 \end{minipage}\\[5em]

\begin{minipage}{0.5\textwidth}
 \begin{fmffile}{diagram2}
\begin{fmfgraph*}(60,30)
\fmfleftn{i}{2}\fmfrightn{o}{4}
\fmflabel{$e^-,p_1$}{i1}\fmflabel{$e^+,p_2$}{i2}
\fmflabel{$e^-,p_3$}{o1}\fmflabel{$e^+,p_4$}{o2}
\fmf{photon,label=$\gamma$}{v5,v2}
\fmf{photon,label=$k_{1}$}{o3,v3}
\fmf{photon,label=      $\  \ \ \ k_{2}$}{o4,v4}
\fmf{fermion}{i1,v5,i2}
\fmf{fermion}{o1,v4,v3,v2}
\fmf{fermion}{v2,o2}
\fmfforce{(0.6w,0.5h)}{v2}
\fmfforce{(0.7w,0.8h)}{v4}
\fmfforce{(0.6w,h)}{o1}
\fmfforce{(0.7w,0.6h)}{v3}
\fmfforce{(0.85w,0.25h)}{o2}
\fmfforce{(0.9w,0.4h)}{o3}
\fmfforce{(0.95w,0.8h)}{o4}
 \end{fmfgraph*}
\end{fmffile}\quad\\
\end{minipage}\\[5em]
 
 \begin{minipage}{0.5\textwidth}
 \begin{fmffile}{diagram3}
\begin{fmfgraph*}(60,30)
\fmfleftn{i}{2}\fmfrightn{o}{4}
\fmflabel{$e^-,p_1$}{i1}\fmflabel{$e^+,p_2$}{i2}
\fmflabel{$e^-,p_3$}{o1}\fmflabel{$e^+,p_4$}{o2}
\fmf{photon,label=$\gamma$}{v5,v2}
\fmf{photon,label=$k_{1}$}{o3,v3}
\fmf{photon,label=      $\  \ \ \ k_{2}$}{o4,v4}
\fmf{fermion}{i1,v5,i2}
\fmf{fermion}{o2,v4,v3,v2}
\fmf{fermion}{v2,o1}
\fmfforce{(0.6w,0.5h)}{v2}
\fmfforce{(0.6w,0.2h)}{v4}
\fmfforce{(0.8w,0.9h)}{o1}
\fmfforce{(0.7w,0.38h)}{v3}
\fmfforce{(0.5w,0h)}{o2}
\fmfforce{(0.9w,0.4h)}{o3}
\fmfforce{(0.85w,0.05h)}{o4}
\end{fmfgraph*}
\end{fmffile}\quad\\
\end{minipage}\\[5em]
 
 \begin{minipage}{0.5\textwidth}
 \begin{fmffile}{venga}
\begin{fmfgraph*}(60,30)
\fmfleftn{i}{4}\fmfrightn{o}{2}
\fmflabel{$e^-,p_1$}{i1}\fmflabel{$e^+,p_2$}{i2}
\fmflabel{$e^-,p_3$}{i3}\fmflabel{$e^+,p_4$}{i4}
\fmf{photon,label=$\gamma$}{v1,v3}
\fmf{photon,label=$k_{1}$}{o1,v2}
\fmf{photon,label=      $\  \ \ \ k_{2}$}{o2,v4}
\fmf{fermion}{i1,v2,v1,i2}
\fmf{fermion}{i3,v4,v3,i4}
\fmfforce{(0.4w,0.15h)}{v2}
\fmfforce{(0.8w,0.15h)}{o1}
\fmfforce{(0w,0.7h)}{i4}
\fmfforce{(0.1w,1h)}{i3}
\fmfforce{(0.4w,0.85h)}{v4}
\fmfforce{(0.8w,0.85h)}{o2}
 \end{fmfgraph*}
\end{fmffile}
\end{minipage}\\[2em]

 \caption{Some of the Feynman diagrams associated to the two-photon decay 
 channel of Ps$_{2}$, Ps$_{2}\rightarrow \gamma \gamma$. There are four more 
 Feynman diagrams that contribute to this decay channel, but they can be obtained 
 as cross terms of the diagrams 
 shown here, {\it i.e.}, with interchanged momenta of the outgoing photons 
 ($k_{1}\leftrightarrow k_{2}$). }
\end{figure}

\noindent 
The quantity $|\Psi_{Ps_{2}}\left( 0,0,0,0 \right)|^{2}$ represents the 
probability of finding the four fermions at the same point in position space. Some details 
about its calculation are given below. Eq. (\ref{eq-1}) 
can be viewed as an extension of the previous generalization 
of Kryuchkov \cite{Kryuchkov-1994} where a three body initial state 
was taken into account for the single photon decay of Ps$^{-}$. $\mathcal{M}$
 represents the transition matrix associated with the decay 
 channel, and therefore $|\mathcal{M}|^{2}$ represents the 
probability for such a transition. It is obtained by averaging the 
squared modulus of the total amplitude $\mathcal{A}$ over the spin states  
of the incoming particles [e$^{-}$($p_{1}$,$s_{1}$),e$^{+}$($p_{2}$,$s_{2}$)
,e$^{-}$($p_{3}$,$s_{3}$),e$^{+}$($p_{4}$,$s_{4}$), here $s_{i}$ represents 
the spin of each particle] and by summing
over the polarizations of the outgoing particles [$\epsilon(k_{1})$, 
$\epsilon({k_{2}})$, here $\epsilon(k_{i})$ denotes the 
polarization of each photon], {\it i.e.},

\begin{equation}
\label{eq-2}
|\mathcal{M}|^{2}=\sum_{\epsilon(k_{1})}\sum_{\epsilon(k_{2})}\frac{1}{2^{4}}\sum_{s_{1}}
\sum_{s_{2}}\sum_{s_{3}}\sum_{s_{4}}|\mathcal{A}|^{2}.
\end{equation}

\noindent
The amplitude $\mathcal{A}$ associated with the decay channel
 Ps$_{2}\rightarrow \gamma \gamma$ contains eight terms, each 
 of them associated with every Feynman diagram that contributes 
 to the process (see Fig.1). The amplitude is given by
 
\begin{eqnarray}
\label{eq-3}
\mathcal{A}=e^{4}\bigg[ \bar{v}(p_{4},s_{4})\gamma^{\lambda} \epsilon_{\lambda}(k_{1})
\frac{\slashed{p}_{4}-\slashed{k}_{2}+m_{e}}{(p_{4}-k_{2})^{2}-m_{e}^{2}}\gamma^{\nu} \nonumber \\
\times \frac{\slashed{p}_{3}-\slashed{k}_{1}+m_{e}}{(p_{3}-k_{1})^{2}-m_{e}^{2}}\gamma^{\sigma}
\epsilon_{\sigma}(k_{1})u(p_{3},s_{3}) \nonumber \\
\times \frac{g_{\mu \nu}}{(p_{1}+p_{2})^{2}}\bar{v}(p_{2},s_{2})\gamma^{\mu}u(p_{1},s_{1})\nonumber \\
+\bar{v}(p_{4},s_{4})\gamma^{\nu} \frac{\slashed{p}_{3}-\slashed{k}_{1}-\slashed{k}_{2}+m_{e}}{(p_{3}-k_{1}-k_{2})^{2}-m_{e}^{2}}
\gamma^{\lambda} \nonumber \\ 
\times \epsilon_{\lambda}(k_{2}) \frac{\slashed{p}_{3}-\slashed{k}_{1}+m_{e}}{(p_{3}-k_{1})^{2}-m_{e}^{2}}\gamma^{\sigma}
\epsilon_{\sigma}(k_{1})u(p_{3},s_{3}) \nonumber \\
\times \frac{g_{\mu\nu}}{(p_{1}+p_{2})^{2}}\bar{v}(p_{2},s_{2})\gamma^{\mu}u(p_{1},s_{1}) \nonumber \\
+\bar{v}(p_{4},s_{4})\gamma^{\sigma}\epsilon_{\sigma}(k_{1}) \frac{\slashed{p}_{4}-\slashed{k}_{1}+m_{e}}{(p_{4}-k_{1})^{2}-m_{e}^{2}}
\gamma^{\lambda} \nonumber \\ 
\times \epsilon_{\lambda}(k_{2}) \frac{\slashed{p}_{4}-\slashed{k}_{1}-\slashed{k}_{2}+m_{e}}{(p_{4}-k_{1}-k_{2})^{2}-m_{e}^{2}}\gamma^{\nu}
u(p_{3},s_{3}) \nonumber \\
\times \frac{g_{\mu\nu}}{(p_{1}+p_{2})^{2}}\bar{v}(p_{2},s_{2})\gamma^{\mu}u(p_{1},s_{1}) \nonumber \\
+\bar{v}(p_{4},s_{4})\gamma^{\lambda}\epsilon_{\lambda}(k_{2}) \frac{\slashed{p}_{4}-\slashed{k}_{2}+m_{e}}{(p_{4}-k_{2})^{2}-m_{e}^{2}}
\gamma^{\nu} \nonumber \\ 
\times u(p_{3},s_{3}) \frac{g_{\mu\nu}}{(p_{2}+p_{1}-k_{1})^{2}} \bar{v}(p_{2},s_{2}) \gamma^{\mu} \nonumber \\
\frac{\slashed{p}_{1}-\slashed{k}_{1}+m_{e}}{(p_{1}-k_{1})^{2}-m_{e}^{2}}\gamma^{\sigma}
\epsilon_{\sigma}(k_{1})u(p_{1},s_{1}) + (k_{1}\leftrightarrow k_{2}) \bigg], \nonumber \\
\end{eqnarray} 

\noindent
Here the Feynman gauge has been employed as well as 
the slashed notation, {\it i.e.}, $\slashed{p}=\gamma^{\nu}p_{\nu}$. 
The $\gamma$ matrices are related with the Dirac matrices as
defined in Ref. \cite{Peskin}. Once the amplitude 
of the process is known $\mathcal{A}$, the transition probability 
associated with the decay channel at hand can be found, 
by means of Eq. (\ref{eq-2}). The calculations needed 
are rather involved, so they have been undertaken using the 
software program Mathematica \cite{Mathematica}, yielding 

\begin{equation}
\label{eq-4}
\Gamma_{Ps_{2}\rightarrow \gamma \gamma}=|\Psi_{Ps_{2}}\left( 0,0,0,0 \right)|^{2}
\frac{521}{512}\frac{\pi^{3}}{2}\frac{\alpha^{4}}{m_{e}^{8}}.
\end{equation}

\begin{figure}[t]
\includegraphics[width=10.cm]{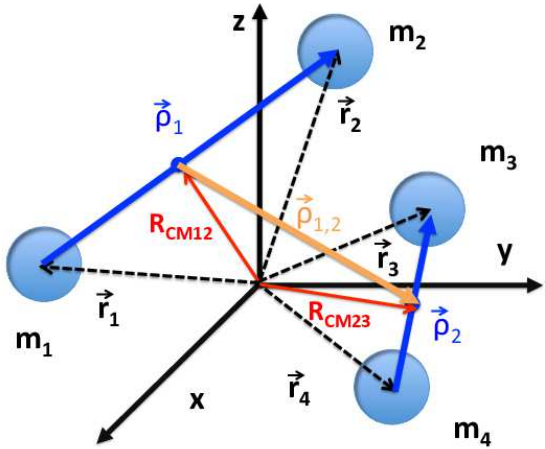}
\caption{Jacobi coordinates for the four-body problem.}
\label{default}
\end{figure}

The Ps$_{2}$ ground state wave function has been obtained by 
using hyperspherical coordinates \cite{Hyperspherical} in conjunction with 
an explicitly correlated gaussian basis, following the 
method of Daily and Greene \cite{Daily-2014}. In 
particular, the Ps$_{2}$ wave function is described 
in terms of the Jacobi coordinates depicted in 
Fig. 2, $\Psi_{Ps_{2}}(\rho_{1},\rho_{2},\rho_{1,2},\rho_{CM})$. 
After neglecting the center of (CM) motion (since the interaction 
potential does not depend on $\rho_{CM}$) and using adiabatic 
hyperspherical approximation, the wave function may be
expressed as $\Psi_{Ps_{2}}(R,\Omega)$, where $R$ denotes 
the hyperradius and $\Omega$ labels the solid angle element associated to 
the eight hyperangles needed for the characterization of a four-body 
collision (neglecting the CM motion). Here the normalization condition for the 
wave function is 

\begin{equation}
\label{eq-5}
\frac{\int |\Psi_{Ps_{2}}(R,0)|^{2}R^{8}dR}{\int d\Omega}=1.
\end{equation}
 
\noindent
Finally, taking into account that $|\Psi_{Ps_{2}}(0,0,0,0)|^{2}=\frac{|\Psi_{Ps_{2}}(0,0)|^{2}}{\sqrt{\Omega}}$, 
one finds $|\Psi_{Ps_{2}}(0,0,0,0)|^{2}$=4.5$\times 10^{-6}$ a$_{0}^{-9}$, 
with a$_{0}$ the Bohr radius. This value is in 
good agreement with the value reported previously by 
Frolov, 4.56 $\times 10^{-6}$ a$_{0}^{-9}$ \cite{Frolov-2009}.

After inserting the probability to find the four fermions 
at the same point $|\Psi_{Ps_{2}}(0,0,0,0)|^{2}$, the relation between atomic 
units and natural units, and after taking into account Eq. (\ref{eq-4}), we find $\Gamma_{Ps_{2}\rightarrow \gamma \gamma}$ = 9.0 $\times 10^{-12}$ s$^{-1}$. 
This decay rate is smaller than the alternative decay channels explored thus far, and which have been
previously reported by Folov \cite{Frolov-2009}. Table I shows a 
comparison between the rate for the two-photon decay and
all the decay channels previously reported. Table I implies 
that the rate reported here, although smaller than the rest, is still
comparable with the zero-photon decay channel. It is related with 
the number of vertices in each decay channel. The zero-photon decay 
involves three vertices ,whereas the two-photon decay channels 
require four vertices. This difference implies that $|\mathcal{M}|^{2}$ 
has an extra factor of $\alpha$ for the case of two-photon decay, in comparison 
with the zero-photon decay.

\begin{table}[h]
\caption{Decay rates for Ps$_{2}$ molecule in (s$^{-1}$). The 
decay rates labelled as $\Gamma_{n\gamma}$, previous calculated 
by Frolov \cite{Frolov-2009}, refers to the annihilation of electron-positron
 pairs in the Ps$_{2}$ molecule, being $n$ the number of photons emitted. 
 Whereas $\Gamma_{Ps_{2}\rightarrow \gamma \gamma}$ stands for 
 the four-body collision among the four fermions leading to the 
 formation of two photons, see text for details. }
\begin{center}
\begin{tabular}{l c c}
\hline
\hline
 Decay Channel &  & Decay rate (s$^{-1}$) \\
 \hline
 $\Gamma_{0\gamma}$ &  & 2.32 $\times 10^{-9}$ \\
 $\Gamma_{1\gamma}$ &  & 1.94 $\times 10^{-1}$ \\
 $\Gamma_{2\gamma}$ &  & 4.44 $\times 10^{9}$ \\
 $\Gamma_{Ps_{2} \rightarrow \gamma\gamma}$ &  & 9.0 $\times 10^{-12}$ \\
 $\Gamma_{3\gamma}$ &  & 1.20 $\times 10^{7}$ \\
 $\Gamma_{4\gamma}$ &  & 6.56 $\times 10^{3}$ \\
 $\Gamma_{5\gamma}$ &  & 0.11 $\times 10^{2}$\\
 \hline
 \hline
 \end{tabular}
\end{center}
\label{default}
\end{table}%

\section{Conclusions}

The two-photon annihilation rate of Ps$_{2}$ has been 
calculated using a non-relativistic reduction of quantum 
electrodynamic methods. This annihilation process 
refers to the simultaneous decay of two electrons and two 
positrons into two photons, providing a rare but 
unambiguously unique signature of the presence of 
the Ps$_{2}$ molecule. All the Feynman diagrams 
contributing to such process have been taken into 
account for the calculation of the transition probability.  
The wave function for ground state Ps$_{2}$ has been 
calculated by employing correlated Gaussian basis 
functions in combination with hyperspherical coordinates \cite{Daily-2014}.
The annihilation rate for this process turns out to be $\Gamma_{Ps_{2}\rightarrow \gamma \gamma}$
=  9.0 $\times 10^{-12}$ s$^{-1}$. 
While this value is smaller than that of other decay channels of Ps$_{2}$, 
it is nevertheless in the same range as the rate associated with the zero-photon 
decay \cite{Frolov-2009}.

The observation of the event studied here will be 
very challenging due to its very long lifetime. However, from 
a fundamental point of view, the two-photon annihilation of Ps$_{2}$ 
constitutes a way to sample the Ps$_{2}$ wave function, from a 
four-body perspective, yielding crucial information about the 
nature of the bound state. Finally, we point out that in some astrophysical regions such as near the galactic center where a high 
density of positrons and electrons are available, this event may 
be observed, due to its unique emission signature of two photons with energies equal to 1.022 MeV. 
This region of the gamma ray spectra remains largely unexplored to date, although 
the International Gamma-Ray Astrophysics Laboratory (INTEGRAL) 
telescope has the capability for it. Indeed this telescope has 
found the signatures of two-photon annihilation in Ps \cite{INTEGRAL}.

\section{Acknowledgements}

The authors would thank K. M. Daily for 
supplying the value of the Ps$_{2}$ wave function 
at the origin. S. T. L. thanks T. Clark for enjoyable discussions. 
This work was supported by the U.S. Department
of Energy, Office of Science, Basic Energy Sciences, under
Award number DE-SC0010545 (for J.P.-R. and C.H.G.).

\bibliography{Ps2}

\end{document}